\documentclass[sigconf]{acmart}
\usepackage{subfig}	
\newcommand{\tabincell}[2]{\begin{tabular}{@{}#1@{}}#2\end{tabular}}
\usepackage{makecell}
\usepackage{setspace}

\AtBeginDocument{%
  \providecommand\BibTeX{{%
    \normalfont B\kern-0.5em{\scshape i\kern-0.25em b}\kern-0.8em\TeX}}}

\copyrightyear{2022} 
\acmYear{2022} 
\setcopyright{acmcopyright}\acmConference[ICPC '22]{30th International Conference on Program Comprehension}{May 16--17, 2022}{Virtual Event, USA}
\acmBooktitle{30th International Conference on Program Comprehension (ICPC '22), May 16--17, 2022, Virtual Event, USA}
\acmPrice{15.00}
\acmDOI{10.1145/3524610.3527923}
\acmISBN{978-1-4503-9298-3/22/05}

\begin{document}

\title{Towards Exploring the Code Reuse from Stack Overflow during Software Development}
\author{Yuan Huang}
\affiliation{%
  \institution{Sun Yat-sen University}
    \institution{School of Software Engineering}
  \city{Zhuhai}
  \country{China}
  \postcode{519082}
}
\email{huangyuan5@mail.sysu.edu.cn}

\author{Furen Xu}
\affiliation{%
  \institution{Sun Yat-sen University}
    \institution{School of Software Engineering}
  \city{Zhuhai}
  \country{China}
  \postcode{519082}
}
\email{xufr@mail2.sysu.edu.cn}

\author{Haojie Zhou}
\affiliation{%
  \institution{Sun Yat-sen University}
    \institution{School of Computer Science and Engineering}
  \city{Guangzhou}
  \country{China}
  \postcode{510006}
}
\email{zhouhj8@mail2.sysu.edu.cn}

\author{Xiangping Chen}
\authornote{Corresponding author.}
\affiliation{%
  \institution{Sun Yat-sen University}
    \institution{School of Communication and Design}
  \city{Guangzhou}
  \country{China}
  \postcode{510006}
}
\email{chenxp8@mail.sysu.edu.cn}

\author{Xiaocong Zhou}
\affiliation{%
  \institution{Sun Yat-sen University}
    \institution{School of Computer Science and Engineering}
  \city{Guangzhou}
  \country{China}
  \postcode{510006}
}
\email{isszxc@mail.sysu.edu.cn}

\author{Tong Wang}
\affiliation{%
  \institution{Sun Yat-sen University}
    \institution{School of Computer Science and Engineering}
  \city{Guangzhou}
  \country{China}
  \postcode{510006}
}
\email{wangtong2@mail2.sysu.edu.cn}


\begin{abstract}
As one of the most well-known programmer Q\&A websites, Stack Overflow (i.e., SO) is serving tens of thousands of developers every day. Previous work has shown that many developers reuse the code snippets on SO when they find an answer (from SO) that functionally matches the programming problem they encounter in their development activities. To study how programmers reuse code on SO during project development, we conduct a comprehensive  empirical study. First, to capture the development activities of programmers, we collect 342,148 modified code snippets in commits from 793 open-source Java projects, and these modified code can reflect the programming problems encountered during development. We also collect the code snippets from 1,355,617 posts on SO. Then, we employ CCFinder to detect the code clone between the modified code from commits and the code from SO, and further analyze the code reuse when programmer solves a programming problem during development. We count the code reuse ratios of the modified code snippets in the commits of each project in different years, the results show that the average code reuse ratio is 6.32\%, and the maximum is 8.38\%. The code reuse ratio in project commits has increased year by year, and the proportion of code reuse in the newly established project is higher than that of old projects. We also find that some projects reuse the code snippets from many years ago. Additionally, we find that experienced developers seem to be more likely to reuse the knowledge on SO. Moreover, we find that the code reuse ratio in bug-related commits (6.67\%) is slightly higher than that of in non-bug-related commits (6.59\%). Furthermore, we also find that the code reuse ratio (14.44\%) in Java class files that have undergone multiple modifications is more than double the overall code reuse ratio (6.32\%). 
\end{abstract}



\keywords{Stack Overflow, Code Reuse, GitHub, Code Clone, Software Development, Code Commit}


\maketitle

\section{Introduction}
\label{sec:introduction}

In recent years, the rise of programming Q\&A websites has promoted the sharing and dissemination of programming knowledge \cite{barua2014developers}. The explosive growth of software data has spawned many innovative applications\cite{huang2019software,liu2014imashup}. How to make full use of these software data for research has become particularly important. Stack Overflow is one of the most well-known programmer Q\&A sites, which has attracted much attention by many developers and researchers since its establishment in 2008. Not only can it help the questioner solve the problem, but also enable the respondent to get satisfaction from altruistic behavior \cite{ma2014knowledge}.

The SO website has become a toolbox for a large number of developers \cite{vasilescu2013stackoverflow}. According to the statistical data from the SO website, as of October 2021, the number of people visiting the website exceeded 100 million per month. It can be seen that SO has brought great help and convenience to many developers. According to our statistics, there are 1,581,814 posts related to Java programming language (cut-off date: 2021/02/28), about 392 posts released per day, which shows that the SO community is very active.

\begin{figure*}[ht]
	\centering
	\includegraphics[width=\textwidth]{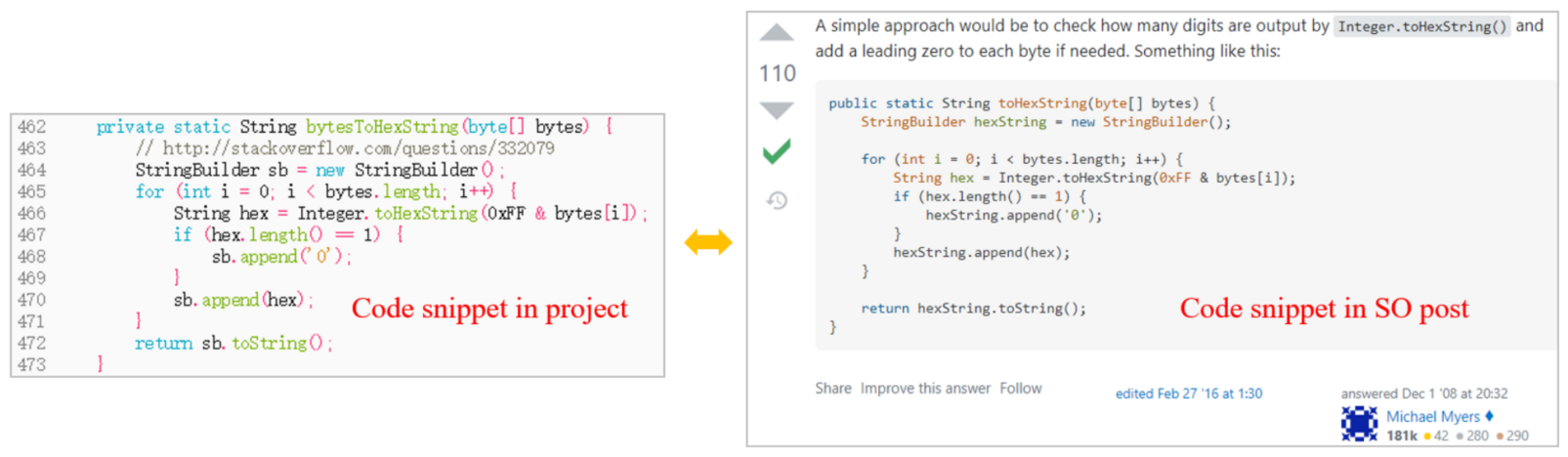}
	\caption{Code reuse example between SO and project}
	\label{fig0}
	\end{figure*}

Previous study \cite{7958574} shows that many developers reuse the code snippets on SO when they find an answer from SO that functionally matches their programming problem. The code reuse from SO can bring great convenience for developers, which also saves their programming time and improve the programming efficiency. An real example\footnote{https://searchcode.com/codesearch/view/43953373/} is shown in Figure \ref{fig0}, where a developer reuses the code from SO when he/she develops a program function, i.e., \begin{ttfamily}check how many digits are output by Integer.toHexString() and add a leading zero to each byte if need\end{ttfamily}.

In this paper, we conduct an empirical study to explore how programmers reuse code on SO during development. To achieve this goal, we need to capture the development process of programmers in a project. The commits from a project may record the bug fixing or new feature addition, and most of them reflect the programming problems encountered during development. Then, we can analyze reuse code during development by analyzing the code reuse between the modified code snippets in the commits and the code snippets on SO.
Therefore, we firstly crawl the 793 most popular java  projects from GitHub, which contain 342,148 modified code snippets in the commits.  At the same time, we collect the question-and-answer pairs  (2008/08/01-2021/02/28) from SO, and obtain 1,355,617 java-related posts after filtering. 

Since the goal of this study is to investigate the code reuse from SO during development, we need to determine the reused pairs of the code between SO and GitHub. Therefore, we employ CCFinder, a token-based clone detection algorithm proposed by Kamiya et al.\cite{kamiya2002ccfinder}, to identify the cloning relationship between the code snippets on SO and the modified code snippets in the commits. For the definition of code reuse from SO, we identify potential code reused pairs based on clone detection results and chronological order (i.e., the code on SO must be published earlier than the code in commit), which is similar to the way used in these studies \cite{abdalkareem2017code,an2017stack}. The reason why not using explicit links as study \cite{wu2019developers} is that: there only a small proportion of code reuse recorded SO links. Since our goal is to analyze the trend of code reuse and the overall situation of code reuse during software development, which needs a large dataset. So using explicit links is not suitable. To facilitate research and application, the replication package and the dataset are available at \url{https://github.com/love-SE/code-reuse-research}.

To gain insights into the practical value of the investigation we make, our survey revolve around the following research questions and get the following results(for further discussion, please refer to the corresponding chapter):

	\textbf{RQ1:} How popular is code reuse  in  development? According to our statistics, among the 793 projects, the average code reuse ratio is 6.32\% for all projects. We also observe that the proportion of code reuse in new projects is higher than that in old projects.
	
	\textbf{RQ2:} Are experienced developers more likely to reuse code from Stack Overflow? From our results, the more experienced contributors of the projects, the more likely they reuse the knowledge from SO. This may indicate that after the developers are familiar with the project, they will better leverage the knowledge on SO to solve the problems encountered in the project.
	
	
	\textbf{RQ3:} Are the modified code snippets involved in the commits related to bug fixing more likely to reuse the code from Stack Overflow? From the results, the code reuse ratio of the bug-related modified code snippets involved in the commits is slightly higher than that of the non bug-related. This shows that developers may turn to SO for answers when they encounter bugs in their programming. 
	
	\textbf{RQ4:} Will the code reused from SO be modified multiple times in development? The results show that 14.44\% of the code reused from SO will be modified multiple time, which indicates that the modified code snippets involved in the commits with code reuse are more likely to be modified multiple times by developers. 
	
	\textbf{RQ5:} What type of Stack Overflow post code are more likely to be reused by developers? From the statistical results, the most frequently reused posts are some popular technologies related to Android, Java web development framework Spring, etc. We find that there is a certain relationship between the type of post and the type of its corresponding reused projects. 

The rest of this paper is summarized as follows. Section \ref{sec3} describes data collection process and clone detection method. Section \ref{sec4} presents the research questions and analyzes the survey results. Section \ref{sec5} introduces the related work of the study. Section \ref{sec6} describes the threats to validity. Section \ref{sec7} concludes this study. 

\section{METHODOLOGY}
\label{sec3}
	We first obtain the Q\&A pair data from the SO website, extract the code snippets from each post, then crawl GitHub popular projects, and obtain all the historical commit files from these projects. Specifically, we adopt the Change Distilling algorithm \cite{kamiya2002ccfinder} to extract the modified code snippets in the commits. After constructing these two code datasets, we leverage the efficient clone detection algorithm CCFinder to identify the code reuse between Q\&A website  and open source projects. The following data sets we collect and research questions are all pertained to Java 
development.
	
	\subsection{Data collection}
	
	\textbf{Constructing Stack Overflow Code Database.} Stack Overflow provides a public RESTful API for accessing the data on the website, but the tool is subject to certain restrictions when used and searched, so we choose StackExchange, the public data dump source of the website, and download the posts (a ".xml" format file) as the data source, and the time span of the Q\&A pairs contain in the file is from August 1, 2008 (the establishment of SO) to February 28, 2021. Each Q\&A pair contains detailed information, such as question title, release date, question tag, etc.
	
	Since the data source file is in XML file format, it needs to be parsed to obtain the information we need. We retain posts related to Java by filtering the data based on the question tag "$<$Java$>$" information contained in each Q\&A pair. After filtering, there are a total of 1,581,814 related posts remaining.
	
	Since some posts are in plain text form and do not involve code information, and our investigation focuses on the existence of reused code, we select the posts with code snippets based on its tag "$<$code$>$ $<$ $\backslash$code $>$" in the post information. At the same time, when running the clone detection algorithm, it is necessary to set the minimum length of the algorithm detection. If the length threshold is set too small, the detection algorithm may recognize common statements such as single-line "if statements" or "for statements" as code clones. Obviously this will affect the accuracy of clone detection. Therefore, we filter out code snippets that are too short in advance in the data filtering stage, instead of filtering at detection time by setting the length threshold parameter of the clone detection tool. After filtering the code whose token length is less than 25 (the threshold choice will be discussed in latter section \ref{RQ1}.), the total number of posts is: 1,355,617. Table \ref{tabl1} shows the amount of post data after each stage of filtering.
	
\begin{table}[htbp]
	\setlength{\abovecaptionskip}{0cm} 
	\caption{Number of SO posts}
	\centering
	\label{tabl1}
	\begin{tabular}{cc}  
		\toprule
		Each Stage  & Posts count \\
		\midrule
		All posts        			& 45, 919, 817 	 	\\
		Posts with '$<$Java$>$' tags                & 1,581,814  		    \\
		Posts Filtered by Length    		    & 1,355,617 		     \\
		\bottomrule
	\end{tabular}
\end{table}

	We use the 1,355,617 posts as the source for constructing the SO code dataset. Each post contains a lot of information, such as "OwnerUserID", "OwnerDisplay", and the original text information in the  Q\&A pairs, and we extract the helpful information for our research, such as the publication time information "CreationDate", the title information "Title", the tag "Tags" and so on.


	
	\textbf{Constructing Open-Source Project Code Database.} We crawl 875 most popular Java projects on Github according to the number of stars of theses projects. The collecting time of the commits is from 2001/1/3 to 2020/12/14. In the project collection, we select projects with the label of Java. After filtering, we find that some Github projects belonged to "study notes" (such as project: "toBeTopJavaer"). The content of these projects are mainly text introductions of knowledge points, and most of the modified content involved in the commits are text revisions, which do not reflect the code changes during project development. Hence this type of project will not be included in our dataset. After filtering out the projects with less commits, we get 793 projects, and we collect the commits of these projects. Noting that, we try to take into account every year after 2008 (the time when SO was established), but the commits of popular projects on GitHub are generally distributed after 2011, so we also crawl the commits data distributed between 2008-2010.
	


	
	To identify the code clone between the code snippets involved in the commits and the code snippets on SO, we need to first extract the code snippets according to each commit. “Code snippet” in this study is the code that involves modification in a commit. Specifically, we leverage the Change Distilling algorithm\cite{fluri2007change} to extract the code snippets involved in the commits. The Change Distilling algorithm is used to extract fine-grained code changes by comparing the abstract syntax trees of the new and old code java files involved in a commit. A fine-grained code change is a code snippet, which may have several successive code lines or a single code line.
	
	
	After processing, we get a total of 98,283 commit files (involving 793 popular GitHub projects). Since each commit may contain multiple modified files, we find that the number of modified java files involved in these 98,283 commits is 342, 148. In the following, we call the modified code snippets in the  commits as "CS-GC"(i.e. the abbreviation of modified code snippets involved in commits). In addition, we call the code snippets involved in SO posts as "CS-SO".

	\subsection{Code Clone Detection Method}
	\label{3.2}
	 Code cloning refers to the existence of identical or similar code fragments between code. There are many reasons for code clones. For example, in order to improve development efficiency, some developers copy the source code of other open source projects and integrate them into their own projects. Specifically, similar to these studies \cite{abdalkareem2017code,an2017stack}, we stipulate that there is a potential code reuse relationship between the code snippets of the post on SO and the modified code snippets in the commits that should meet two conditions: these two code snippets should be detected by the code clone detection algorithm as a clone pair and the time of the latter should be later than the time of the former. So we call the code pair that meets the above two conditions as code reuse pair.

	Clone types can be divided into the following 4 types \cite{roy2009comparison}:
	\begin{itemize}
	\item [1)] Type-I: The two code snippets are the same except for spaces, comments and typesetting;
	
	\item [2)] Type-II: The two code snippets are the same except for the names of some identifiers;
	
	\item [3)] Type-III: The difference between the two code snippets is: some sentences are added, deleted or reordered;
	
	\item [4)] Type-IV: The two code snippets have similar functions, but are implemented by different structural variants.
	\end{itemize}

	Some previous research results show that there is a considerable part of the code in the software system (usually 9\% to 17\%)
	consists of cloned code\cite{zibran2011analyzing} and the proportion of code cloned in the code base ranges from 5\% to 50\% \cite{rieger2004insights}. One statistical survey\cite{lopes2017dejavu} of the open source community GitHub shows that 70\% of the code on GitHub consists of clones of previously created files. Their analysis shows that in 9\% to 31\% of projects, at least 80\% code files can be found elsewhere.
	
	The code on SO usually exist in the form of snippets, for example, the respondent usually does not write detailed variable declarations and inheritance relationships for their own code when posting a answer. Therefore, it is difficult to detect by cloning detection algorithms based on syntax, such as some complex methods based on abstract syntax trees\cite{baxter1998clone},\cite{tairas2006phoenix} or program dependency graphs\cite{komondoor2001using},\cite{zou2020ccgraph}. Moreover, our data volume is relatively large, and with some learning-based methods\cite{guo2020graphcodebert},\cite{yu2019neural}, the processing speed will be very inefficient. For comprehensive considerations, we adopt a token-based clone detection algorithm: CCFinder\cite{kamiya2002ccfinder}, which can only detect type I and type II clone types. Wu et al. \cite{wu2019developers}'s survey indicates that when programmers reuse the code on SO, 52\% of code reuse is directly copied, pasted or simply modified, which also shows the rationality of using CCFinder clone detection algorithm.

	CCFinder first converts the code into a form that is easy to handle. CCFinder first remove the initial indentation of each line of each input source code, and add spaces between identifiers and punctuation for processing. Then it replace all the identifiers with a single token character (such as "\$p") and finally compare the sequence after the parameter replacement.
	
\begin{figure}[ht]
	\centering
	\subfloat[]{\includegraphics[width=0.5\linewidth, height=1in]{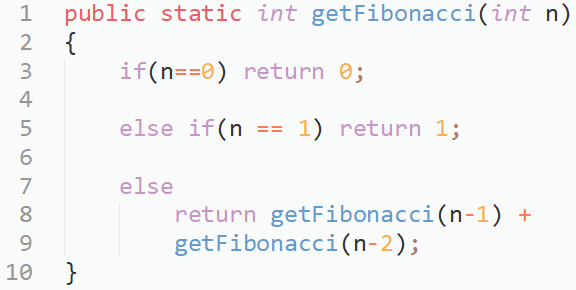}}
	\subfloat[]{\includegraphics[width=0.5\linewidth, height=1in]{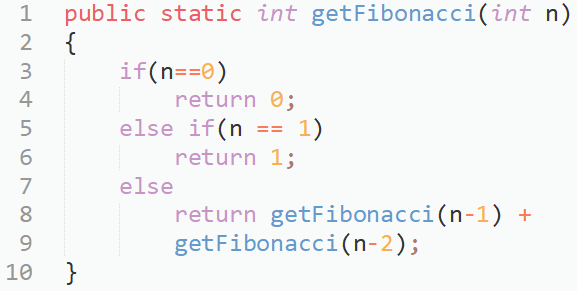}}
	\caption{Type-I code clone pair}
	\label{fun1}
	\end{figure}

\begin{figure}[ht]
	\centering
	\subfloat[]{\includegraphics[width=0.5\linewidth, height=1in]{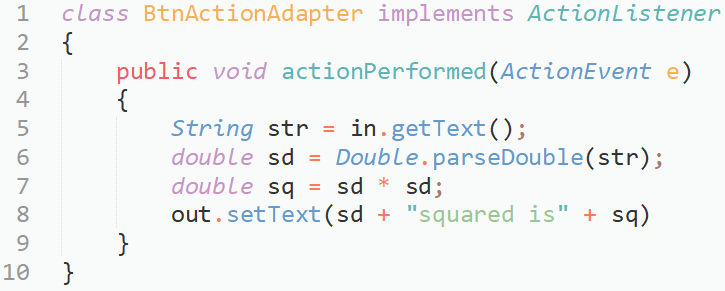}}
	\subfloat[]{\includegraphics[width=0.5\linewidth, height=1in]{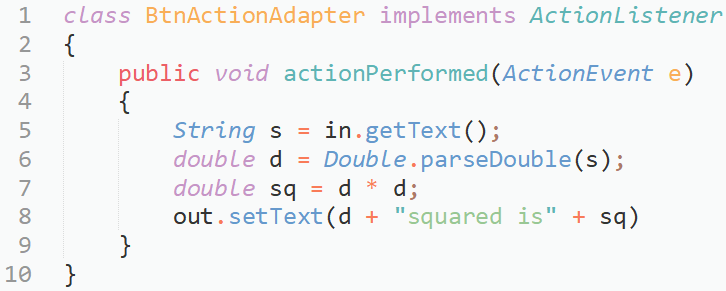}}
	\caption{Type-II code clone pair}
	\label{fun2}
	\end{figure}

	In order to better understand the types of clones that CCFinder can detect, here are two examples, namely type-I and type-II clones. The code snippets in Figure \ref{fun1} can be successfully detected by CCFinder because they only have some differences between spaces and blank lines. Although some variable names have been modified by programmers in Figure \ref{fun2}, they can be successfully detected by CCFinder after preprocessing conversion.
	
\section{EXPERIMENT}
\label{sec4}
	\label{sec4}
	In this section, we present the analysis process and results for the following research questions:
	
	\subsection{(RQ1) How popular is code reuse  in  development?}
	\label{RQ1}
	\textbf{Motivation}. The purpose of our study is to explore how likely it is to reuse the knowledge on SO during the iterative development of the project. Furthermore, we want to explore whether with the rise of Q\&A websites, the code reuse ratio will change over the year. And whether newly created projects will have a higher code reuse ratio than projects built later. Namely, we want to analyze whether code reuse changes over time.
	
	\textbf{Approach}. In order to study how likely it is to reuse the knowledge on SO in the iterative development process of the project, we will input the code snippets involved in the commits and the code snippets of SO into the CCFinder for code clone detection. In the process of extracting the code snippets from SO posts, we filter out the code snippets with less than 10 lines after extraction. Because longer lines of code are conducive to containing the current contextual information of the code, the clone detection of the code will be more accurate than the short code snippets as input. Counting all the code snippets after extracting, we find that the average number of tokens in the snippet is 25, so we filter code snippets in our dataset that are below this length threshold. The parameters of CCFinder are configured according to the default parameters.
	
	\begin{figure}[ht]
	\centering
	\includegraphics[width=\linewidth]{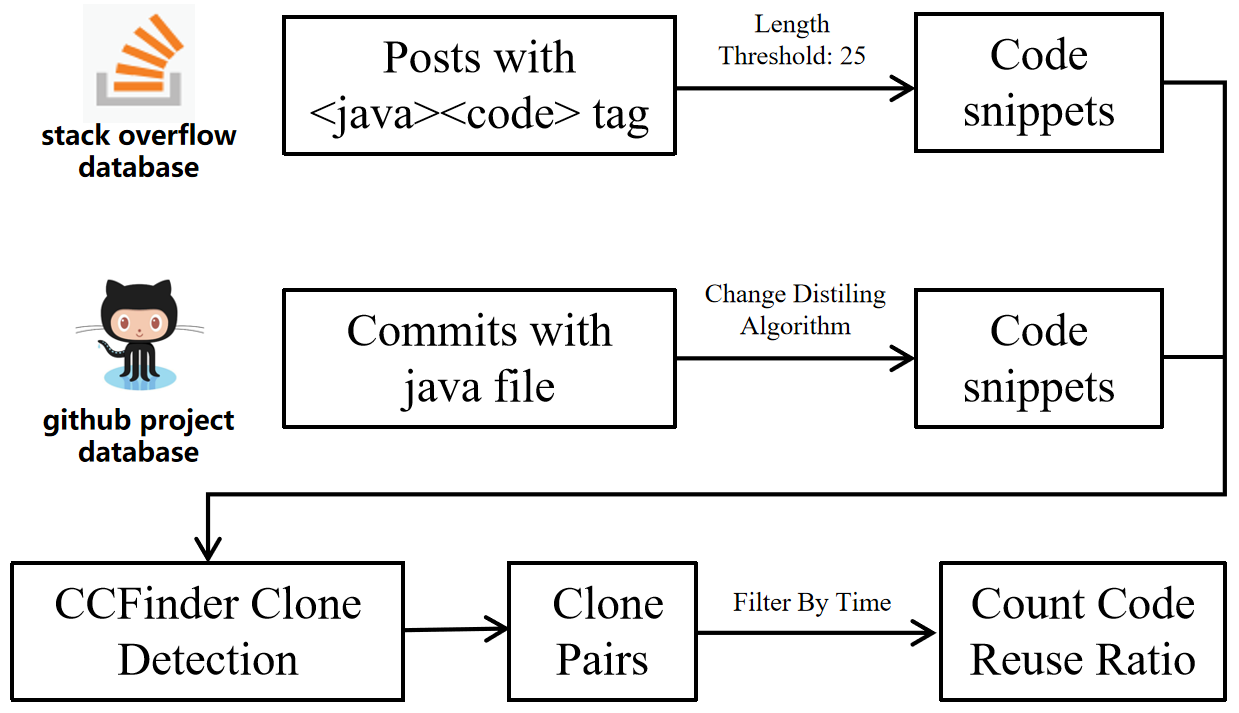}
	\caption{An overview of our approach of code reuse detection}
	\label{fig2}
	\end{figure}
	
	After finishing the clone detection by CCFinder, we can get all the clone pairs. However, we need to filter out some clone pairs that don't meet the time limit. For each clone pair, we require that released time of the commit should be later than that of the post on SO, so that the modified code snippets in the commits may reuse the code from SO. In a real development scenario, reusing the knowledge on SO from the future is obviously unreasonable.
	
	In addition, we find that many cloned code snippets exist in the form of "one-to-many". A certain code snippet involved in the commits may have clone relationship with multiple SO code snippets. This is because the same snippet appears in the answers to multiple questions, or the answers to the same question are sometimes relatively similar. In view of this situation, we select the longest matching as the mapping result. In the absence of a similarity percentage, it is generally believed that the more overlapping token segments in the two pieces of code, the higher the similarity.
	
	For the exploration of whether the code reuse ratio of project will change over the year, we calculate the code reuse ratio of each project in different years based on the clone detection results in the previous step, and then summarize to obtain the average clone ratio of all projects in each year.
	
	As for the investigation of whether newly established projects are more inclined to reuse the knowledge on SO, we use the year when each project first appeared and submit the modification information as the establishment time of the project. Because there are new projects established in 2001-2020, we call the projects that started to update earlier as old projects, and those that started to update later are called new projects. We divide the years into 4 levels of new and old. Specifically, we classify projects established in 2001-2005 as older projects, projects established in 2006-2010 as old projects, and projects established in 2011-2015 as new projects, projects established in 2016-2020 are called as newer projects.
	
	For the investigation of code reuse transitivity, that is, some projects may reuse the code from many years ago and then the new projects may reuse the code from these projects, which will lead to a large time difference between CS-GC and CS-SO with code reuse relationship. We first select CS-GC and CS-SO with code reuse relationship, and then count the time difference of each code reuse pair. Specifically, since a CS-GC may have multiple CS-SO reuse candidates in the detection result, for the time difference statistics, we only select the latest reused CS-SO.
	
\begin{figure}[ht]
		\centering
		\includegraphics[width=\linewidth]{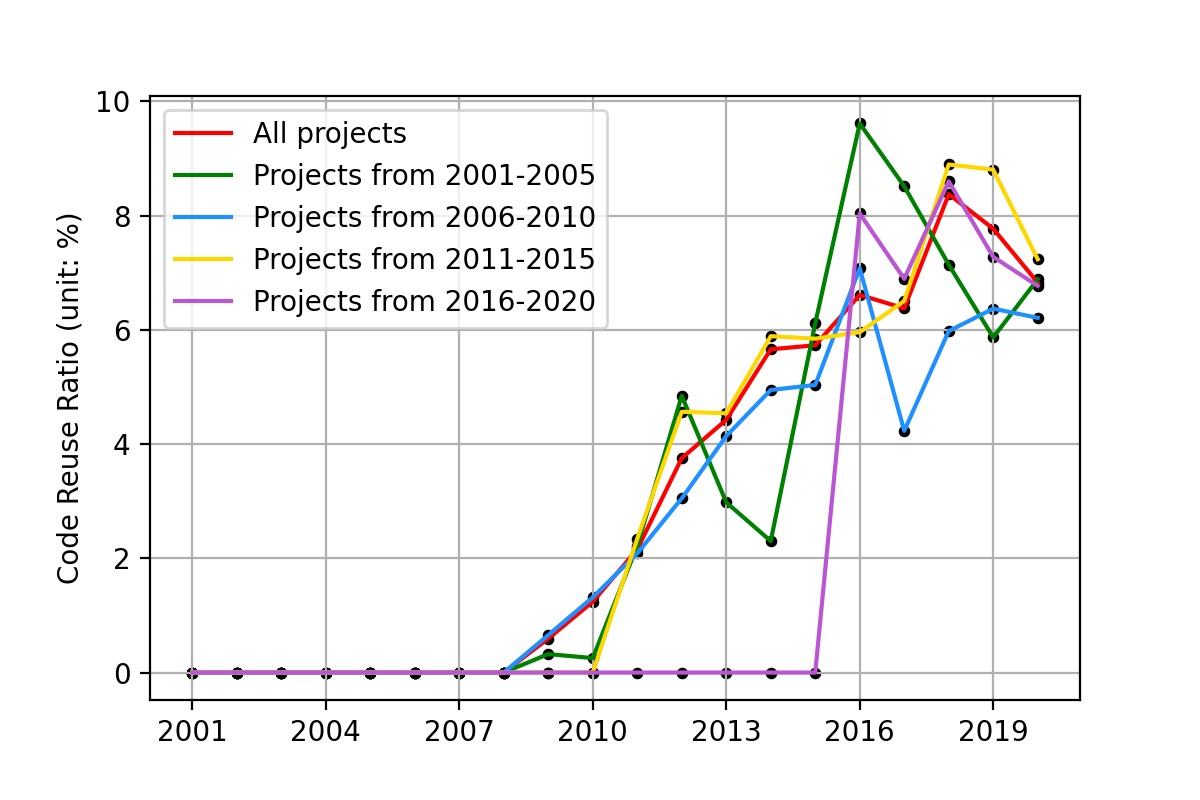}
		\caption{The code reuse ratio with its trend over the years}
		\label{fig3}
	\end{figure}

	\textbf{Results}. According to statistics, the code reuse ratio of each project in open-source code database ranges from 0.59\% to 8.38\%. If taking all projects as a whole, the average code reuse ratio is 6.32\%, which proves that developers will reuse the knowledge on the SO Q\&A website during the iterative development of the project. From the red curve in Figure \ref{fig3} , we can see that from the overall trend, the code reuse ratio is increasing year by year, reaching a maximum of 8.38\% in 2008. In addition, there will be a small decline in 2019 and 2020. Perhaps due to the rise of the open source movement in recent years, developers may reuse the code from some open source code websites such as GitHub, etc. Even with a small decline, the overall trend is still rising. This further proves that with the rise of the SO Q\&A website in recent years, developers increasingly tend to reuse the knowledge on the SO Q\&A website.
	
	At the same time, considering that if only the total project code reuse ratio changes over time, it is very likely that the general trend is caused by newly established projects. In order to explore whether the code reuse ratio in the old and newly established projects is also increasing year by year, we separately count the old projects (i.e., the establishment time is 2001-2005 and 2006-2010) and the newly established projects (i.e., the establishment time is 2011-2015 and 2016-2020). Obviously from the overall trend in Figure \ref{fig3} , the code reuse ratio of the old and newly established projects is increasing year by year, and the code reuse ratio of the newly established projects (the yellow curve and the purple curve) is basically higher than that of the old ones (the green curve and the blue curve), indicating that the newly established projects will more likely reuse the knowledge on SO.

	Specifically, we divide 4 levels of new and old according to the year when the project was established. The number of projects at each level is shown in Table \ref{tab2} . It can be seen that the establishment time of most of the 793 projects is distributed between 2010-2020.
	\begin{table}[ht]
		\setlength{\abovecaptionskip}{0cm} 
		\centering
		\caption{Number of projects with different levels of old and new}
		\begin{tabular}{cc}  
			\toprule
			Old\&New Level  & Project count \\
			\midrule
			Older (2001-2005)        		& 6 \\
			Old (2006-2010)   & 51 \\
			New (2011-2015)   & 372 \\
			Newer (2016-2020)        		& 364 \\
			\bottomrule
		\end{tabular}
		\label{tab2}
	\end{table}

\begin{figure}[ht]
	\centering
	\includegraphics[width=\linewidth]{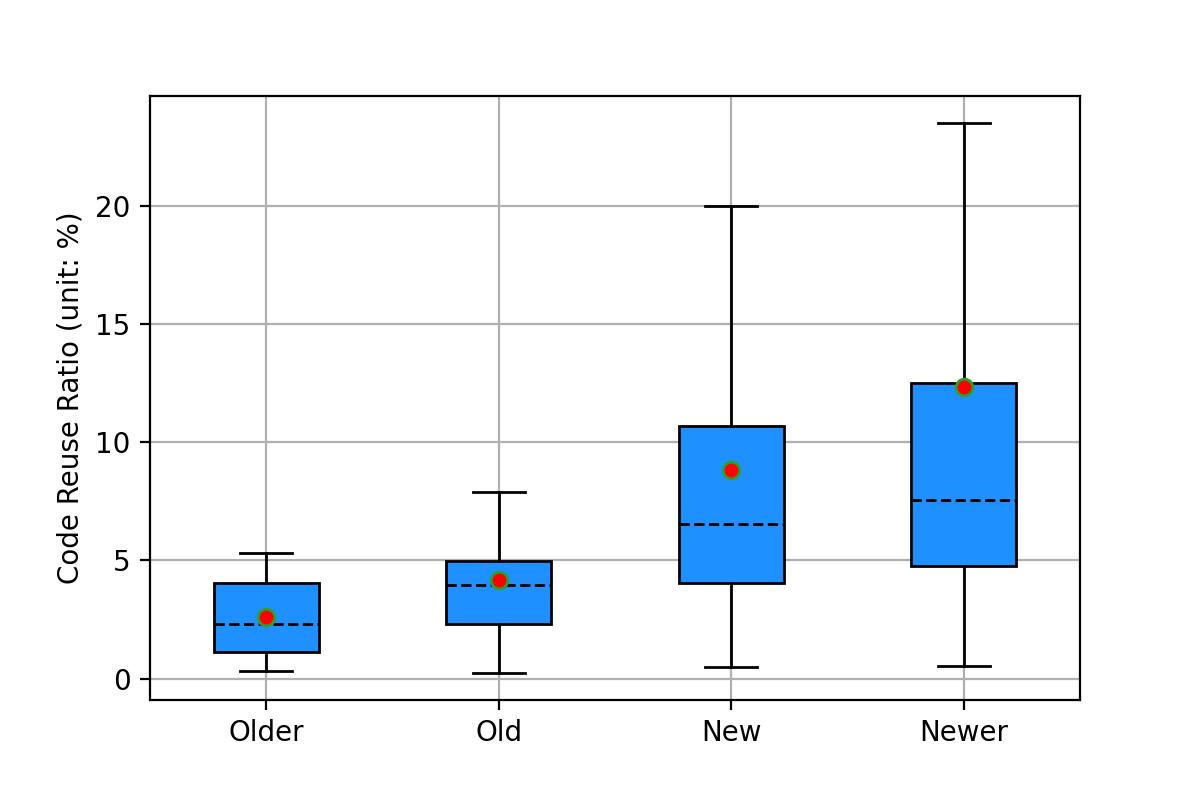}
	\caption{Box plot of code reuse ratios for 4 levels of projects}
	\label{fig4}
	\end{figure}

	The distribution of code reuse ratio in these 4 new and old levels of projects is shown in Figure \ref{fig4}. In addition, there are some projects with a code reuse ratio close to 0, because the project development time is relatively long, and the main commits are distributed in the early stage of the project. At this time, the SO website has not yet emerged, so there are few similar code between the two. The purpose of this statistics is mainly to study the relationship between the establishment time of new and old projects and the code reuse ratio, so our following statistics do not include this part of the project. From the average of the following statistical results, the code reuse ratio is 2.60\%, 4.15\%, 8.81\%, 12.34\%, and the newer the project, the higher the code reuse ratio. And judging from the median of the statistical results, the same is true. Among them, it can be clearly seen that the code reuse ratio of new projects at level 4 is sufficiently higher than the code reuse ratio of old projects at level 1 by 9.74\%. This verifies that the code reuse ratio in newly established projects is higher than the proportion of code reuse in old projects. It also shows that developers are more inclined to reuse the knowledge on the SO Q\&A website in the development of newly established projects.
	
\begin{figure}[ht]
	\centering
	\includegraphics[width=\linewidth]{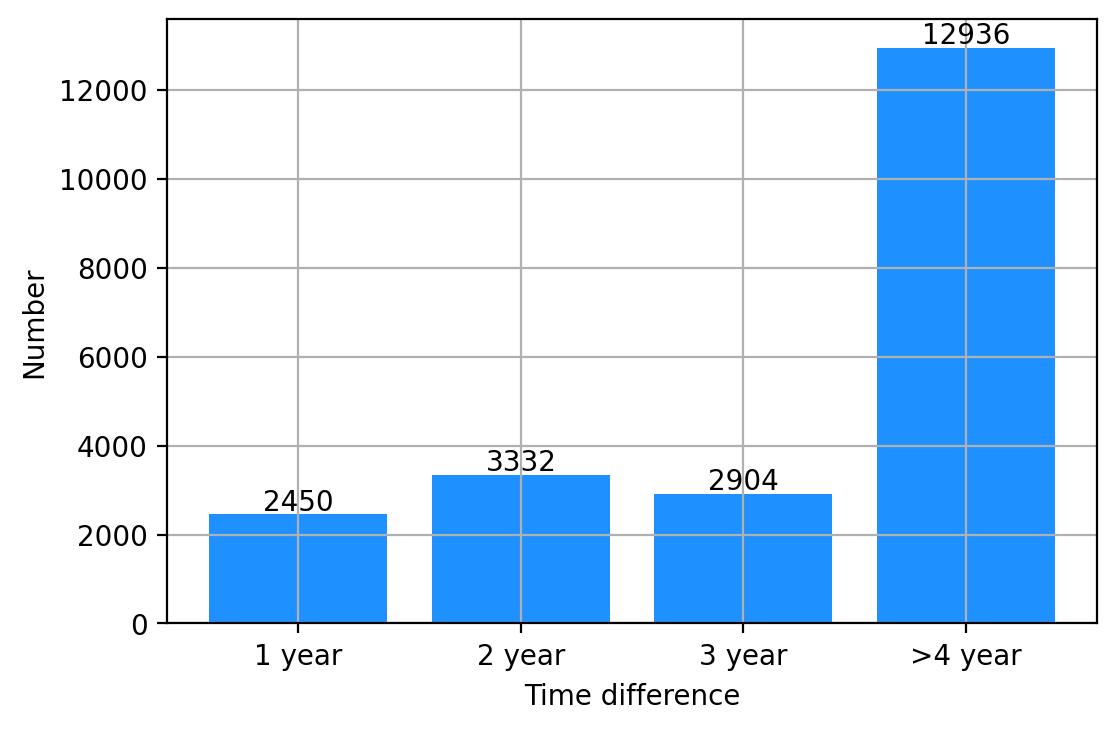}
	\caption{Distribution of time difference between CS-GC and CS-SO with code reuse relationship}
	\label{time_diff}
	\end{figure}

	In Figure \ref{time_diff}, from the time difference between CS-SO and CS-GC with code reuse relationship, it can be seen that the time difference of more than 4 years accounts for a high proportion, which may explain the transitivity of code cloning to a certain extent, that is, some projects may reuse the code from many years ago and then the new projects may reuse the code from these projects, which will lead to a large time difference between the modified code snippets in the commits and code snippets in SO posts with code reuse relationship.
	
	In general, from the above analysis results, code reuse activities may become more and more popular in software development. It is inevitable that a code snippet 
    coming from SO may be reused in same or different projects for multiple times. Then, if the code snippet from SO is defective, the defect may be propagated to multiple projects. Therefore, locating the symmetrical vulnerabilities in same or different 
    projects is a worthy research direction for researchers.

	\subsection{(RQ2) Are experienced developers more likely to reuse the code from Stack Overflow?}
	\textbf{Motivation}. When developing a project, some experienced developers may be better at retrieving the programming knowledge they need on SO for reusing because they have a good grasp of some professional knowledge, and some novices may not have enough knowledge of some programming knowledge or professional terminology, so their ability to retrieve programming knowledge may be relatively limited. For this motivation, we launch this research question.
	
	\textbf{Approach}. We measure the committer’s experience level based on the number of commits they contributed to the project. We assume that the more commits developer contribute, the higher the project experience they have. We count the proportion of all committers with different experiences who reuse the programming knowledge on SO. In addition, we also count the code reuse ratio in the commit of committers with different experiences.
	
	\textbf{Results}. According to statistics, we find that most of the project contributors’ commits are within 100, so we divide them into six types of committers in Table \ref{tab3} with different project experience levels. From Table \ref{tab3} , we can clearly see that the more experienced committers are, the higher the proportion of them who have reused the knowledge on SO. This may indicate that after the developers are familiar with the project, they will better leverage the knowledge on SO to solve the problems encountered in the project. Additionally, we recommend that SO platform should advertise its site to novice programmers.

	\begin{table}[ht]
		\setlength{\abovecaptionskip}{0cm} 
		\centering
		\caption{Code reuse ratio of programmers with different levels of experience(from the perspective of the committer)}
		\begin{tabular}{ccccc}  
		\toprule
		\tabincell{c} {Interval}&\tabincell{c} {Commit\\count} &\tabincell{c}{Committer\\count} &\tabincell{c}{Committer \\count\\(reuse SO)} &Ratio\\
		\midrule
		1 &1	&2,630	&165	&6.27$ \% $ \\
		2 &2	&1,275	&140	&10.98$ \% $ \\
		3 &3-5	&1,776	&341	&19.20$ \% $ \\
		4 &6-15	&1,773	&694	&39.14$ \% $ \\
		5 &16-100	&1,659	&1,205	&72.63$ \% $ \\
		6 &$>$100	&657	&633	&96.34$ \% $ \\
		\bottomrule
		\end{tabular}
		\label{tab3}
	\end{table}

Furthermore, we count the code reuse ratio from the perspective of the modified code snippets involved in the commits, it can be seen from the Table \ref{tabcom} , the number of CS-GC that reused the knowledge from SO in the interval 4 is slightly higher than other intervals. This may be because developers with moderate experience are more familiar with the project than novices, and they are good at reusing the knowledge on SO. Compared to more experienced developers, they may need to reuse the knowledge on SO even more.
	
	\begin{table}[ht]
	\setlength{\abovecaptionskip}{0cm} 
	\centering
	\caption{Code reuse ratio of programmers with different levels of experience(from the perspective of CS-GC)}
	\begin{tabular}{ccccc}  
		\toprule
		\tabincell{c} {Interval}&\tabincell{c} {Commit\\count} &\tabincell{c}{CS-GC\\count} &\tabincell{c}{CS-GC count\\(reuse SO)} &Ratio\\
		\midrule
		1 &1	&2,630	&165	&6.27$ \% $ \\
		2 &2	&2,550	&160	&6.27$ \% $ \\
		3 &3-5	&6,721	&452	&6.72$ \% $ \\
		4 &6-15	&16,656	&1,179	&7.07$ \% $ \\
		5 &16-100	&63,157	&4180	&6.61$ \% $ \\
		6 &$>$100	&250,434	&15,486	&6.18$ \% $ \\
		\bottomrule
	\end{tabular}
	\label{tabcom}
\end{table}

	\subsection{(RQ3) Are the modified code snippets involved in the commits related to bug fixing more likely to reuse the code from Stack Overflow?}
	\textbf{Motivation}. Considering that when some developers encounter bug-related issues in developing projects, they may go to some Q\&A websites, such as the SO platform, to seek help or find relevant knowledge. For non-bug-related issues, developers may solve them more by themselves instead of reusing to the knowledge of Q\&A websites. For this motivation, we conduct RQ3 to verify whether bug-related code snippets involved in the commits are more likely to reuse the knowledge on SO.
	
	\textbf{Approach}. In the open-source project code database we build, each CS-GC has a corresponding commit message. We distinguish whether the commit is bug-related or not based on the bug-related vocabulary. Among them, bug-related vocabulary (because some of the annotation information is in Chinese, we also include bug-related vocabulary in Chinese): bug, fix, solve, issue, problem, error, repair, defect, vulnerable, vulnerability. Because we match based on strings, other parts of speech like fix, such as fixing, fixes, will be matched. After dividing the commits into bug-related commits and non-bug-related commits, we count the number of the bug-related CS-GC and the non-bug-related CS-GC that reuse the code snippets on SO based on the clone detection results of the CCFinder algorithm and the chronological order of the SO post and the commit.
	
	\textbf{Results}. From the results in Table \ref{tab5}, the overall difference in the proportion of bug-related and non-bug-related CS-GC with reusing the knowledge on SO is not distinguished, and the proportion of the former will be slightly higher than that of the latter, which can suggest that developers should not reuse code directly from SO, but check its security risk. This also shows that developers do not only reuse the knowledge from SO when they encounter bugs. For example, when developers make some modifications that are not related to bugs: adding new features or optimizing code structure, they may also reuse the knowledge on SO.

	\begin{table}[ht]
	\renewcommand\tabcolsep{0pt} 
	\setlength{\abovecaptionskip}{0cm} 
	\centering
	\caption{Statistics of code reuse ratio of bug-related and non-bug-related CS-GC}
	\begin{tabular}{cccc}  
		\toprule
		&\tabincell{c}{Bug-related\\CS-GC}	&\tabincell{c}{Non-bug-related\\CS-GC}	&\tabincell{c}{All CS-GC}\\
		\midrule
		CS-GC count	&57,252	&284,896	&342,148\\
		\tabincell{c}{CS-GC(reuse SO)\\count}	&3,636	&17,986	&21,622 \\
		\tabincell{c}{Code Reuse Ratio}	&6.35$ \% $	&6.31$ \% $	&6.32$ \% $\\
		\bottomrule
	\end{tabular}
	\label{tab5}
	\end{table}

	\subsection{(RQ4) Will the code reused from SO be modified multiple times in development?}
	\textbf{Motivation}. In the process of developing projects, developers may reuse the code from open source projects or Q\&A websites. These reused codes may have certain security risks, which causes developers to make multiple modifications for maintenance purposes. Therefore, based on this motivation, we want to investigate whether the reused code snippets involved in the commits are more likely to be modified multiple time (i.e. from this perspective, we investigate whether the reused code snippets are more likely to have security risks).
	
	\textbf{Approach}. We first separately count the multiple modifications between all modified java files involved in the commits of each project, and then perform clone detection of the modified code snippets that have multiple modifications with the code snippets of the SO code base. Specifically, we use the CCFinder clone detection algorithm to detect similar modified commit code snippets in the project, and then determine whether the code snippets involved in the commits belong to the same java class in the same project, and if the conditions are met, it is considered to be multiple modifications of the same java class. Furthermore, we divide the class files that have been modified multiple times into two categories (i.e., those with reusing the SO posts and those without reusing the SO posts), and then respectively count the code reuse ratios corresponding to the class files with different modification times.
	
	In the next step, we input CS-GC and CS-SO into the CCFinder algorithm for code clone detection.

	\textbf{Results}. The experimental results in Table \ref{tab7} show that the code reuse ratio (14.44\%) in the modified java files involved in the commits that has undergone multiple modifications (i.e. the number of modifications is more than twice) is significantly higher than that in the overall code reuse ratio(6.32\%). From the perspective of higher modification times, the code reuse ratio has not decreased, but basically maintained at 12\%-15\%, which shows that the java files involved in the commits with code reuse is more likely to be modified by developers for multiple times.

	\begin{table}[ht]
	\renewcommand\tabcolsep{5pt} 
	\setlength{\abovecaptionskip}{0cm} 
	\centering
	\caption{Code reuse ratio corresponding to different modification times}
	
	\begin{tabular}{cccc} 
	\toprule
	\tabincell{c}{Modification\\time} &\tabincell{c}{Commit\\count}	&\tabincell{c}{Commit\\(reuse SO)\\count}&\tabincell{c}{Code Reuse\\Ratio} \\
	\midrule
	1	&309,778	&16,948	&5.47$ \% $ \\
	2, 3	&13,536	&2,137	&15.78$ \% $ \\
	4, 5	&9,026	&1,168	&12.94$ \% $ \\
	6, 7	&4,322	&622	&14.39$ \% $ \\
	8, 9	&2,283	&322	&13.26$ \% $ \\
	$ \ge $10	&3,203	&425	&14.10$ \% $ \\

	\bottomrule
	\end{tabular}
	\label{tab7}
	\end{table}
	
\begin{figure}[ht]
	\centering
	\includegraphics[width=\linewidth]{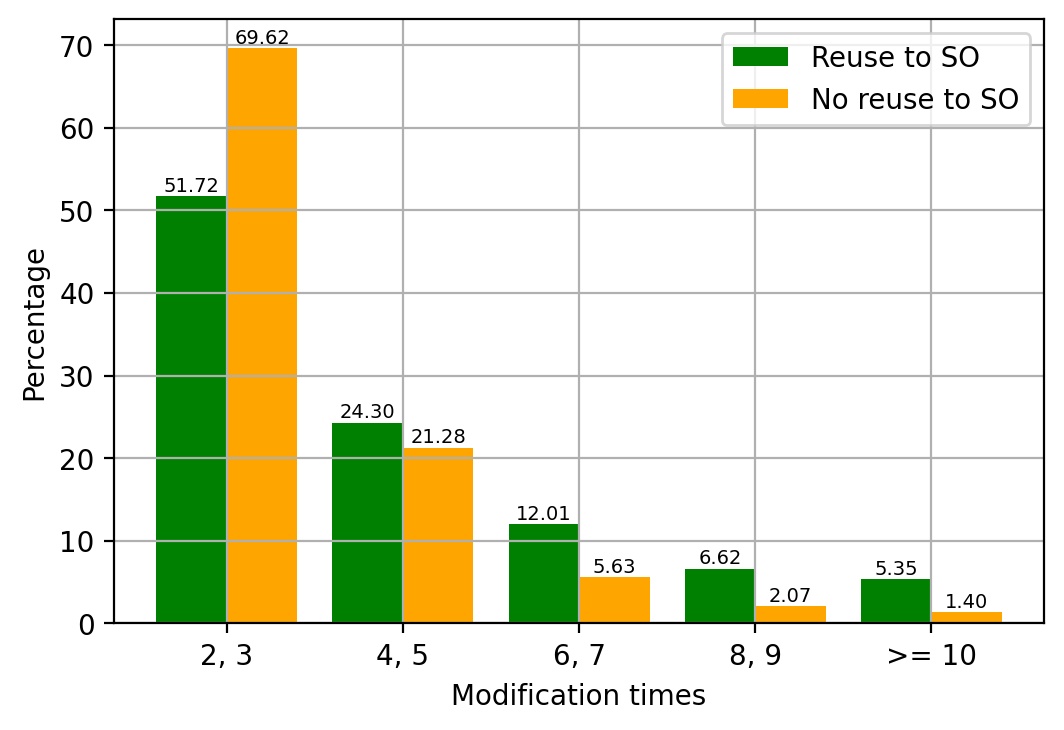}
	\caption{The distribution of java classes involved in the commits that reuse CS-SO and not reuse CS-SO}
	\label{multiple modification}
	\end{figure}

	It can be seen from the comparison results in the Figure \ref{multiple modification} that the proportion of classes with reusing the SO posts is higher than that without reusing the SO posts(except for those with modification times of 2 and 3). To some extent, this may reveal the security risks brought by code reuse, which leads to more modifications. Therefore, developers should pay more attention to code reuse.
	
	\subsection{(RQ5) What type of Stack Overflow posts are more likely to be reused by developers?}
	\textbf{Motivation}. Since we are investigating posts with the tag $<$Java$>$ on SO, we further want to investigate what type of SO posts are more likely to be reused by developers. At the same time, we want to analyze in details the distribution of different types of posts reused by different projects, and further analyze what knowledge is mainly involved in certain types of posts.
	
	\textbf{Approach}. Specifically, we count the number of each post in the SO code base based on the clone detection results obtained by inputting the SO code base and the open-source project code base into the CCFinder algorithm. We consider posts that have been reused more than twice as re-reused posts, and then based on the tag information of these re-reused posts, we can calculate what type of these posts belong to. In addition, we analyze the proportion of the modified code snippets involved in the commits that reuse a tag-related post from the project granularity. Furthermore, we conduct an in-depth analysis of posts related to a certain tag.
	
\begin{figure}[ht]
	\centering
	\includegraphics[width=\linewidth]{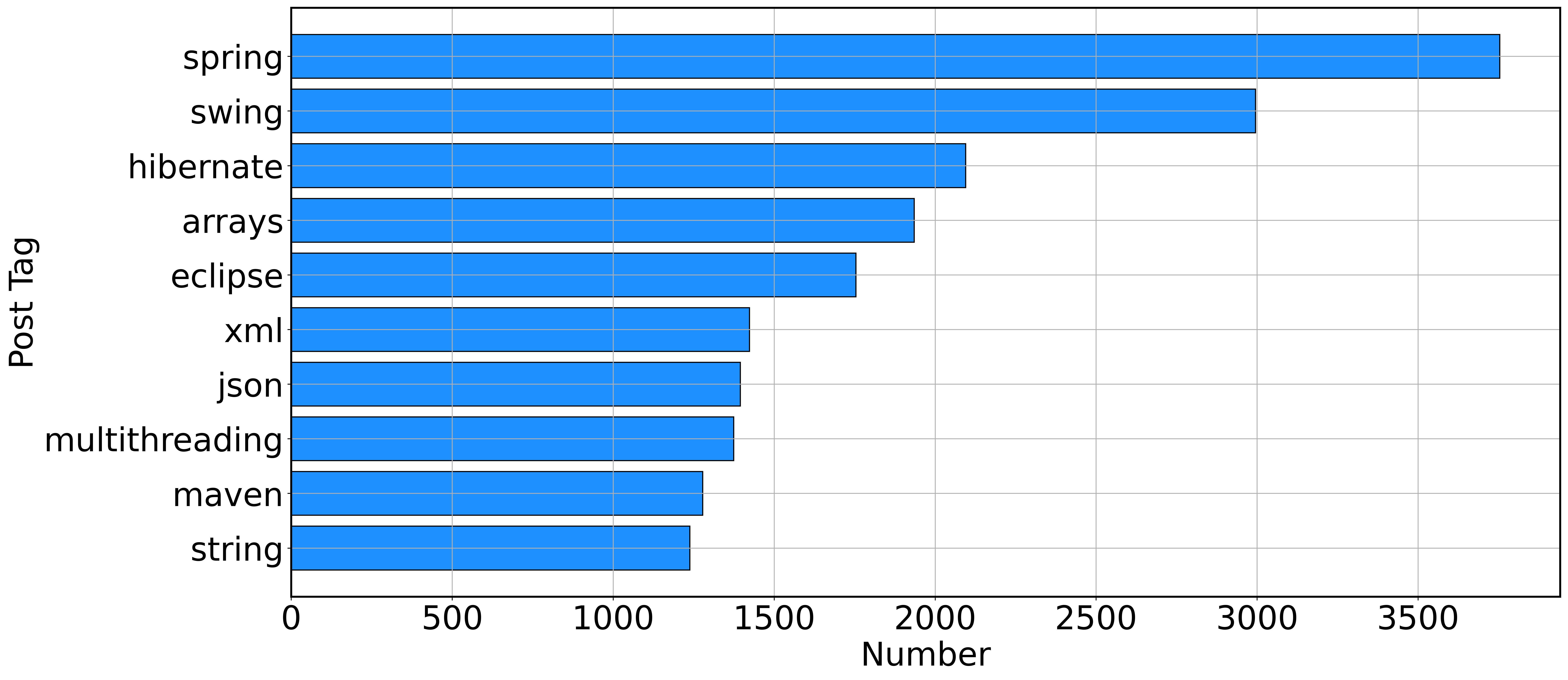}
	\caption{Post tags of top 10 reuse}
	\label{fig6}
	\end{figure}

	\textbf{Results}. According to statistics, the number of posts involved in the SO code base is 1,355,617. Among them, the number of reused posts is 77,599, accounting for 5.72\% of the total number of posts. Among the 77,599 posts that are reused, the number of re-reused posts is 61,126, accounting for 78.77\% of them. This shows that most of the posts on SO that are reused will be re-reused. We filter out the posts that have been repeatedly reused, and then count the number of tags of each type in the tag information, and display the top 10 tags by the histogram, as shown in Figure \ref{fig6} (all the tags involved are visualized). The larger the size of the word in the word cloud graph, the larger the proportion of posts with that tag(Since the filtered posts are all posts with the $<$Java$>$ tag on SO, the $<$Java$>$ tag is not taken into consideration here). Furthermore, we separately count the GitHub projects involved in the posts corresponding to these 10 tags. As shown in Table \ref{tab8} , the meaning of the data in the table is: for example, in the first row of data, the second column indicates the number of GitHub projects involved in posts with the tag of $ < $android$ > $. The third column of data indicates that the GitHub project reuses to the top 5 data of the number of $ < $android$ > $ related posts. The number in parentheses after the project name indicates the number of the $ < $android$ > $ posts this project reused.

	Based on the above statistical results, we can see from Figure \ref{fig6} that the most frequently reused posts are some popular technologies related to Android, Java web development framework Spring, and Java GUI toolkit Swing, etc. From Table \ref{tab8} , we can see that there is a certain relationship between the type of post and the type of its corresponding projects that reuse it. For example, most of the projects that reuse the android-related posts are related to android, and most of the projects that reuse the spring-related posts are related to spring. And some projects may reuse the posts related to a variety of popular technologies, such as projects FirebaseUI-Android, OpenHub, etc. Additionally, The results show that posts with different tags have different reuse frequency. Therefore, SO platform should consider to improve the way of organizing posts according to their tags, so that developers can quickly find answers.
	
	\begin{table}[ht]
		\renewcommand\tabcolsep{0pt} 
		\setlength{\abovecaptionskip}{0cm} 
		\centering
		\caption{Distribution of 5 popular projects and its corresponding post tags}
		\begin{spacing}{0.1}

			\begin{tabular}{ccc}  
				
				\toprule
				
				\makecell[c]{Tag}& \makecell[c]{Project Count} & \makecell[c]{Project(Top5)}\\
				\midrule
				android        & 492		& \tabincell{c}{FirebaseUI-Android(1213),\\ LifeHelper(1071),\\DoraemonKit(1001),\\ MPAndroidChart(999),\\xxl-job(997)} \\
				\hline
				\specialrule{0em}{1pt}{0pt}
				spring & 430  &  \tabincell{c}{FirebaseUI-Android(431), \\DoraemonKit(372),\\ LifeHelper(369),\\ xxl-job(365), \\spring-boot(317)} \\
				\hline
				\specialrule{0em}{1pt}{0pt}
				swing  & 437 &  \tabincell{c}{FirebaseUI-Android(370), \\MPAndroidChart(347),\\ LifeHelper(323),\\ OpenHub(316),\\ Phonograph(302)} \\
				\hline
				\specialrule{0em}{1pt}{0pt}
				hibernate & 424  &  \tabincell{c}{FirebaseUI-Android(247),\\ DoraemonKit(218), \\LifeHelper(214), \\xxl-job(211),\\ OpenHub(202)} \\
				\hline
				\specialrule{0em}{1pt}{0pt}
				arrays & 413  &  \tabincell{c}{FirebaseUI-Android(240), \\LifeHelper(202), \\MPAndroidChart(197),\\ DoraemonKit(197), \\xxl-job(195)} \\
				\hline
				\specialrule{0em}{1pt}{0pt}
				eclipse  & 410 &  \tabincell{c}{MPAndroidChart(217), \\FirebaseUI-Android(217), \\LifeHelper(209), \\OpenHub(199), \\android-advancedrecyclerview(191)} \\
				\hline
				\specialrule{0em}{1pt}{0pt}
				xml  & 390 &  \tabincell{c}{FirebaseUI-Android(159), \\DoraemonKit(152), \\xxl-job(151), \\LifeHelper(131),\\ MPAndroidChart(129)} \\
				\hline
				\specialrule{0em}{1pt}{0pt}
				json & 386 &  \tabincell{c}{FirebaseUI-Android(188),\\ LifeHelper(164),\\ DoraemonKit(149),\\ xxl-job(147), \\OpenHub(141)} \\
				\hline
				\specialrule{0em}{1pt}{0pt}
				\tabincell{c}{multi-\\threading} & 395  &  \tabincell{c}{FirebaseUI-Android(169),\\ LifeHelper(157),\\ MPAndroidChart(157), \\DoraemonKit(153),\\ xxl-job(143)} \\
				\hline
				\specialrule{0em}{1pt}{0pt}
				maven  & 368 &  \tabincell{c}{FirebaseUI-Android(169), \\LifeHelper(147),\\ OpenHub(132),\\ android-advancedrecyclerview(129),\\ DoraemonKit(129)} \\
				\bottomrule
			\end{tabular}
		\end{spacing}
		\label{tab8}
	\end{table}
	
	In particular, we analyze the posts with the $<$eclipse$>$ tag and selected the 10 most reused posts related to $<$eclipse$>$ tags. We can see that most of these posts are related to the configuration and use of eclipse, as well as the configuration of dependency libraries project needs, etc.

	We select 5 popular projects, and then calculate the percentage of posts they reuse, as shown in the Table \ref{prj2tag} . For example, FirebaseUI-Android is an open source project for Android that allows you to quickly connect common UI elements to Firebase APIs like the Realtime Database or Firebase Authentication. From the top 10 tags involved in the project, it can be seen that most of tags are related to the project, such as android-related posts and UI-related swing-related posts. Judging from the tags involved in the other four projects, most of them may reuse some of the more popular technologies.

	\begin{table}[ht]
	\renewcommand\tabcolsep{2pt} 
	\setlength{\abovecaptionskip}{0cm} 
	\centering
	\caption{Distribution of posts reused by 5 popular projects}
	\begin{spacing}{0.1}
		
		\begin{tabular}{cccc}  
			
			\toprule
			\makecell[c]{Project}& \makecell[c]{Tags\\count} & \makecell[c]{Tag types\\count} & \makecell[c]{Top10 tags}\\
			\midrule
			\tabincell{c}{FirebaseUI\\Android}        & 23,585	&2,992	& \tabincell{c}{android(5\%), spring(2\%),\\
				swing(2\%), hibernate(1\%),\\ arrays(1\%), eclipse(1\%),\\
				json(1\%), maven(1\%),\\ multithreading(1\%), xml(1\%)} \\
			
			\hline
			\specialrule{0em}{1pt}{0pt}
			LifeHelper & 20,532  &2,744 & \tabincell{c}{android(5\%), spring(2\%),\\ swing(2\%), hibernate(1\%),\\ eclipse(1\%), arrays(1\%),\\ 
				json(1\%), multithreadin(1\%),\\ maven(1\%), string(1\%)} \\
			\hline
	
			\specialrule{0em}{1pt}{0pt}
			spring-boot  & 13,309  &2,248 &  \tabincell{c}{android(5\%), spring(3\%),\\spring-boot(2\%), swing(1\%),\\ arrays(1\%),hibernate(1\%),\\json(1\%),maven(1\%),\\javafx(1\%), xml(1\%)} \\
			\hline
					
			\specialrule{0em}{1pt}{0pt}
			OpenHub  & 18,895  &2,602 &  \tabincell{c}{android(5\%), swing(2\%),\\spring(2\%), hibernate(1\%),\\eclipse(1\%), arrays(1\%),\\json(1\%),multithreading(1\%),\\maven(1\%), string(1\%)} \\
			\hline
			
			\specialrule{0em}{1pt}{0pt}
			 xxl-job & 20,172  &2,820 &  \tabincell{c}{android(5\%), spring(2\%),\\swing(1\%), hibernate(1\%),\\arrays(1\%), eclipse(1\%),\\xml(1\%), multithreading(1\%),\\json(1\%),spring-boot(1\%)} \\

			\bottomrule
		\end{tabular}
	\end{spacing}
	\label{prj2tag}
	\end{table}

\section{RELATED WORK}
\label{sec5}
	We discuss two categories of related work. The first is some empirical studies or investigation on SO and the second is some code reuse research on SO.

	\subsection*{A. Empirical Studies on Stack Overflow}
	Over the past years, there have been a large number of researchers conducting various studies on SO\cite{hsieh2010pay, jan2017analyzing, chen2019reliable, toth2020will, zhang2019empirical, meng2018secure, mamykina2011design, ahmad2019toward, tang2021using, cummaudo2020interpreting, zerouali2021identifying, yin2018poster, wang2018users, yazdaninia2021characterization, yang2016security, bosu2013building, rahman2016empirical, zhang2018code, greco2018stackintheflow}.   
	Specifically, some research focuses on analyzing the design and function of the SO system, and making suggestions for improvement to the SO community. Specifically, Wang et al. \cite{wang2018users} conducted a study on the badge system of SO, they analyzed 3,871,966 revision records that are collected from 2,377,692 SO answers. They found that some users performed a much larger than usual revisions on the badge-awarding days compared to normal days; 25\% of the users did not make any more revisions once they received their first revision-related badge. And some prior studies showed that incentive systems may not always drive certain users in a positive way on Q\&A websites \cite{hsieh2010pay}, \cite{jan2017analyzing}. Tóth et al. \cite{toth2020will} analyzed the reason why some questions on SO were closed. In order to help the users compose good quality questions, they introduced a set of classifiers for the categorization of SO posts prior to their actual submission.
	
	Some research discusses security issues on SO. Zhang et al. \cite{zhang2019empirical} thought that SO has accumulated a lot of software programming knowledge, but over time, some knowledge in the post  may be outdated. If these outdated answers are not processed in time, they may mislead some programmers who reuse the outdated answers and cause security problems. Their results showed that more than half of the obsolete answers were probably already obsolete when they were first posted. These outdated posts may introduce vulnerabilities to programmers who reuse the post. Meng et al. \cite{meng2018secure} conducted an empirical study on SO posts, they found that programming challenges are usually related to APIs or libraries, and also reported various vulnerable coding suggestions.
	
	Some researches comprehensively analyze the reasons for the success of SO. Mamykina et al. \cite{mamykina2011design} found that over 92\% of SO questions about expert topics are answered - in a median time of 11 minutes. They used a mixed methods that combines data analysis with user to explain why SO is successful in expert knowledge sharing.
		
	Different from these prior work, we focus on quantitatively and qualitatively investigating the code reuse activities between SO and GitHub projects. 
	 
	\subsection*{B. Code Reuse Related to Stack Overflow}
	Several studies have been done to investigate the code reuse activities on SO\cite{baltes2019usage},\cite{abdalkareem2017code},\cite{an2017stack},\cite{ragkhitwetsagul2019toxic,wu2019developers,manes2021studying,chen2015crowd,zhang2017detecting,yang2017stack,lotter2018code,wang2020duplicate,baltes2017attribution}. Specifically, the following studies are similar to us. Abdalkareem et al. \cite{abdalkareem2017code} performed an exploratory study focusing on code reuse from SO in the context of mobile apps. Specifically, they investigated how much, why, when, and who reuses code. An et al. \cite{an2017stack} conducted a case study with 399 Android apps, aiming to investigate whether developers respect license terms when reusing code from SO posts. Their results  revealed that developers may have copied the code of apps that were potentially reused from SO to answer SO questions. They desired to raise the awareness of the software engineering community about potential unethical code reuse activities. In addition, Ragkhitwetsagul et al. \cite{ragkhitwetsagul2019toxic} believe that copying code snippets from public Q\&A sites (such as SO) is commonplace during software project development. And this approach may introduce security risks, such as violation of permissions, error propagation, and introduction of vulnerabilities, which will reduce the quality of the project code. And their research found that 72,635 Java-related SO code segments and 111 Java projects are clone pairs in the database they constructed. They selected 2,289 candidate clone pairs, and found that 153 were copied directly from the Qualitas project to SO, and 100 of them were outdated, 10 were bug-hazardous, and 214 code snippets violated the original software license. Wu at al. \cite{wu2019developers} conduct a study on 289 files from 182 projects, while we study on 98,283 commits from 793 projects. Our main difference is that we focus more on the code reuse in the dynamic process of software development, which also can be seen from our paper title:
    "Towards Exploring the Code Reuse from SO during Software Development ".
	
	In comparison, our research is more focused on the code reuse activities during the project development. Specifically, we utilize the CCFinder code clone detection tool to extract the cloning relationship between the code snippets on SO and the modified code snippets involved in the commits, and then determine their code reuse relationship according to the chronological order. Another difference is that we collect 1,355,617 posts from SO and 793 GitHub projects. Instead of focusing on analyzing a specific type of project, the 793 GitHub projects we collect contains various kinds of fields. Furthermore, our study not only focus on the latest version of the project, but on the iterative process of the project. Our paper mainly investigate these research points in terms of (1) the code reuse ratio in open-source projects and its trend over the years, (2) the relationship between developer experience and code reuse, (3) the relationship between code reuse and bug-related modified code snippets involved in the commits.(4) the influence caused by code reuse, (5) the type of SO posts that developers generally like to reuse. 

\section{THREATS TO VALIDITY}
\label{sec6}
	\textbf{External Validity. }In our research with 793 GitHub Java-related projects, we only collect data from the SO platform for research. The average code reuse ratio from 2008 to 2020 is 6.32\%, and the maximum is 8.32\%. If we collect data from other Q\&A websites for analysis, then the code reuse ratio will undoubtedly be higher than our statistics. In the future, we will expand our database to other Q\&A websites. The construction of our database has good scalability, when expanding to other Q\&A websites, we only need to consider the difference in the structure of the knowledge source. Future work is necessary to generalize to other programming language for investigating whether our results hold for other language. Considering that CCFinder is suitable for COBOL, C, C++, Csharp, Java, PlainText and VisualBasic languages, and has good scalability, it is very convenient to leverage CCFinder to extend the clone detection to other programming languages.

	\textbf{Internal Validity. }Firstly, for the definition of code reuse, similar to previous related work \cite{abdalkareem2017code,an2017stack}, we identify potential code reused pairs based on clone detection results and chronological order. Since confirming code reuse relationship from the SO to the GitHub project is a very difficult task, we just use this method to determine the potential code reuse relationship. 
	
	Secondly, the code clone detection algorithm we use is the CCfinder proposed by Kamiya et al. \cite{kamiya2002ccfinder}, which can only detect type I and type II clone types. However, the clone detection types mainly include four types, among which Type-I, Type-II and Type-III belong to syntactic cloning, and Type-IV belongs to semantic cloning. Although our research only detect two types of clones, Wu et al. \cite{wu2019developers}'s survey indicated that when programmers reuse the code on SO, 52\% of code reuse is directly copied, pasted or simply modified, which also shows the rationality of using CCFinder clone detection tool. Therefore, if we consider the clone types of Type III and Type IV, there is no doubt that the proportion of code reuse will be higher than our statistics. In the future, we plan to leverage clone detection tools that can detect type-III and type-IV clone types for more precise detection.

\section{CONCLUSION \& FUTURE WORK}                                                                                                                                                                                                                                  
\label{sec7}
	In this paper, we conduct a comprehensive and in-depth empirical study of code reuse activities between SO and GitHub projects. Specifically, we build SO code database (contains 1,355,617 posts) and Open-source Java project code database (contains 342,148 modified code snippets involved in the commits of 793 projects) respectively. We use CCFinder as our clone detection tool to complete our investigation and analysis. Through the investigation, we get the most important results of our study as shown below:
	\begin{itemize}

	\item [1)] Our exploratory study shows that among the GitHub projects we collect, the average code reuse ratio of different projects in different years is 6.32\%, and the maximum is 8.38\%. Since the SO Q\&A website was established in 2008, the code reuse ratio in GitHub project evolution has increased year by year, and the proportion of code reuse in newly established projects will be higher than that in old projects. Furthermore, we find that older code snippets may be continuously transferred as programmers reuse with each other.
	
	\item [2)] Experienced developers seem to be more likely to reuse the knowledge on SO.
	
	\item [3)] The code reuse ratio in the bug-related modified code snippets  (6.35\%) is slightly higher than the code reuse ratio in non-bug-related modified code snippets involved in the commits(6.31\%)
	
	\item [4)] We also find that the code reuse ratio (14.44\%) in Java class files that have undergone multiple modifications is more than double the overall code reuse ratio (6.32\%). This may also explain to a certain extent the multiple modifications caused by the insecure factors brought about by code reuse.
	
	\item [5)] We statistically analyze the types of posts that are reused more on SO, and analyze the distribution of the types of posts reused from the project granularity. The results show that some of the more popular Java-related technologies (e.g. android, swing, etc.) will be reused more.
	\end{itemize}

	From the results of our empirical research, developers may reuse the programming knowledge on the online Q\&A website when developing projects. In the future, we would like to expand our analysis objects to more Q\&A website platforms and more open-source projects with other programming languages, and design a similar code detection based method, aiming to recommend programming knowledge for developers to provide more convenient learning conditions.

\begin{acks}
This work was supported by the Key-Area Research and Development Program of Guangdong Province of China under Grant No. 2020B010164002, the National Natural Science Foundation of China under Grant (No.61902441, 61902105, 61976061), Guangdong Basic and Applied Basic Research Foundation under Grant No.2020A1515010973. 
\end{acks}




\bibliographystyle{ACM-Reference-Format}
\bibliography{stackoverflow}

\appendix

\end{document}